\documentstyle[prc,aps,psfig]{revtex}
\begin{document}
\twocolumn

\title{\Large\bf On the two-proton emission of $^{45}$Fe - a new type of radioactivity}

\author{ 
J. Giovinazzo, B. Blank, M. Chartier{\footnotemark}, S. Czajkowski, A. Fleury, M.J. Lopez Jimenez{\footnotemark}, M.S.~Pravikoff, J.-C. Thomas} 

\address{
CEN~Bordeaux-Gradignan, Le Haut-Vigneau, F-33175 Gradignan Cedex, France}

\author{ 
F. de Oliveira Santos, M. Lewitowicz, V. Maslov{\footnotemark}, M. Stanoiu}

\address{
Grand Acc\'el\'erateur National d'Ions Lourds, B.P. 5027, F-14076 Caen Cedex, 
France}

\author{ 
R. Grzywacz{\footnotemark}, M. Pf\"utzner }

\address{
Institute of Experimental Physics, University of Warsaw, PL-00-681 Warsaw, 
Poland }

\author{ 
C. Borcea}

\address{
IAP, Bucharest-Magurele, P.O. Box MG6, Rumania}

\author{ 
B.A. Brown}

\address{
Department of Physics and Astronomy and National Superconducting Cyclotron Laboratory, Michigan State University, East Lansing, Michigan 48824-1321, USA}

\maketitle

\begin{abstract}
In an experiment at the SISSI-LISE3 facility of GANIL, the decay of the proton drip-line nucleus $^{45}$Fe has been 
studied after projectile fragmentation of a $^{58}$Ni primary beam at 75 MeV/nucleon impinging on a natural nickel 
target. Fragment-implantation events have been correlated with radioactive decay events in a 16$\times$16 
pixel silicon strip detector on an event-by-event basis. The decay-energy spectrum of $^{45}$Fe 
implants shows a distinct peak at (1.06$\pm$0.04)~MeV with a half-life of $T_{1/2}$~= (4.7$^{+3.4}_{-1.4}$)~ms. 
None of the events in this peak is in coincidence with $\beta$ particles which 
were searched for in a detector next to the implantation detector. For a longer correlation interval, daughter decays 
of the two-proton daughter $^{43}$Cr can be observed after $^{45}$Fe implantation. The decay 
energy for $^{45}$Fe agrees nicely with several theoretical predictions for two-proton 
emission. Barrier-penetration calculations slightly favour a di-proton emission picture over an emission 
of two individual protons and point thus to a $^{2}$He emission mode.  
\end{abstract}

\vspace*{0.2cm}
\vspace*{0.2cm}

\renewcommand{\footnotesep}{0cm}
\renewcommand{\footnoterule}{\rule{0cm}{0cm}\vspace{0cm}}
\renewcommand{\thefootnote}{\fnsymbol{footnote}}
\footnotetext[1]{present address: Oliver Lodge Laboratory, Department of Physics, University of Liverpool, Liverpool, L69 7ZE, UK}
\footnotetext[2]{present address: CEA/DPTA/SPN Bruyères-le-Châtel, F-91680 Bruyères-le-Châtel, France}
\footnotetext[3]{present address: Flerov Laboratory of Nuclear Reactions, Joint Institute for Nuclear Research,
        141980 Dubna, Russia}
\footnotetext[4]{present address: Physics Division, ORNL, Oak Ridge, TN 37831-6371,USA}

An ensemble of protons and neutrons can form a nucleus stable against any radioactive decay only if a subtle 
equilibrium between the number of protons and neutrons is respected.  If this equilibrium 
condition is not respected in a nucleus, it becomes radioactive. For small deviations from equilibrium, the nuclei 
decay by $\beta$ decay. 
The limits of stability, the drip lines, are reached if the nuclear forces are no longer able to bind an ensemble of 
nucleons with a too large neutron or proton excess. Whereas for heavy nuclei, e.g.
$\alpha$ decay, $^{14}$C radioactivity or fission occurs, unstable proton-rich nuclei may decay by emission of 
one proton for odd-Z nuclei or of two protons for even-Z nuclei from their ground 
states. The one-proton ground-state decay has been observed for the first time at GSI Darmstadt in 
1981~\cite{hofmann82,klepper82}. Meanwhile almost 30 cases of proton radioactivity have been identified 
for odd-Z nuclei beyond the proton drip line between Z=51 and Z=83~\cite{woods97} allowing e.g. to establish the 
sequence of shell-model single-particle levels beyond the proton drip line.

Two-proton (2p) radioactivity is predicted since 1960~\cite{goldanskii60} to occur for even-Z proton-rich nuclei 
beyond the drip line. Due to the pairing energy, the 2p candidates cannot 
decay by a sequential emission of two protons as the one-proton daughter is energetically not accessible. 
Therefore, only a simultaneous two-proton emission is possible which can take place in two different ways: i) by 
an isotropic emission of the two protons which then have no angular correlation, i.e. they fill the whole phase 
space available, but in order to easily penetrate through the Coulomb and centrifugal barrier of the daughter 
nucleus they share most probably equally the decay energy; ii) by a correlated emission where in the decay 
the $^{2}$He resonance is formed which decays either already under the Coulomb and centrifugal barrier or outside 
the barrier. In both cases, a zero energy difference between the two protons is most likely. However, for a 
$^{2}$He emission, a small relative angle between the two protons might be observable.

Recent theoretical work~\cite{brown91,ormand96,cole96} showed that $^{45}$Fe, 
$^{48}$Ni, and $^{54}$Zn are the best candidates for two-proton ground-state decay as their 2p Q values are about 
1.1-1.8~MeV, whereas one-proton emission is either energetically forbidden or extremely disfavoured due to 
small one-proton decay energies and very narrow intermediate states. 

For light nuclei, the Coulomb and centrifugal barriers for proton emission are smaller and the width of the 
ground states of these nuclei become large. Thus, one-proton emission is energetically possible through the tails 
of broad intermediate states even for even-Z nuclei. This has been observed experimentally for  
$^{6}$Be~\cite{bochkarev92} and $^{12}$O~\cite{azhari98}. In both cases, the emission pattern is compatible with a 
simultaneous uncorrelated decay into the available phase space. Two-proton emission from excited states has 
been observed in eight cases after $\beta$ decay (see~\cite{fynbo00} for a recent reference) as well as after 
inelastic excitation of $^{14}$O~\cite{bain96} and of $^{18}$Ne~\cite{gomez01}. In all cases, the experimental 
data are either not precise enough to indicate a detailed decay pattern or compatible with a sequential or 
three-body picture with half-lives as short as $10^{-21}$s.

With the advent of projectile-fragmentation facilities equipped with powerful separators for in-flight isotope 
separation, medium-mass proton drip-line nuclei came into experimental reach and two of the above mentioned 
most promising candidates, $^{45}$Fe~\cite{blank96fe45} and $^{48}$Ni~\cite{blank00ni48}, could be observed. 
In addition, part of the 
decay strength of $^{45}$Fe was observed~\cite{giovinazzo01} in the experiment designed to identify for the first 
time $^{48}$Ni~\cite{blank00ni48}. However, due to the set-up of the electronics in this experiment, two-proton 
events triggered the data acquisition only with a very low probability (five possible events at about 
1.1~MeV~\cite{giovinazzo01}).

\begin{figure}[htt]
\centerline{
\psfig{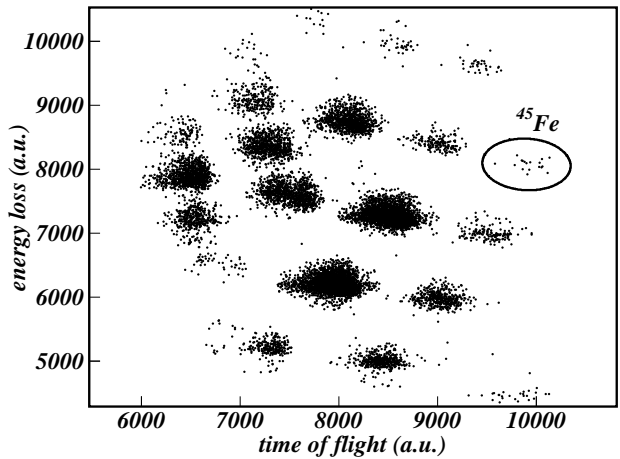}}
\caption{Two-dimensional fragment identification spectrum for the present experiment. The implantation events
         are plotted as a function of their time-of-flight between one channel-plate detector and the silicon stack
         and their energy loss in the first silicon detector. The figure shows only part of the data.}
\protect\label{fe45_id}
\end{figure}

In an experiment performed in June/July 2000 at the SISSI-LISE3 facility of GANIL, we used the projectile fragmentation 
of a $^{58}$Ni primary beam at 75~MeV/nucleon to produce proton-rich nuclei in the range Z=20-28. After production 
in a $^{nat}$Ni target (240 $\mu$m in thickness) located in the SISSI device, the fragments of interest were 
selected by the Alpha/LISE3 separator equipped with an intermediate beryllium degrader (50$\mu$m). 
At the focus of the LISE3 separator, a set-up was mounted to identify (see figure~\ref{fe45_id}) 
and stop the fragments as well as to study their radioactive decays. 
This set-up consisted in two channel-plate detection systems for timing 
purposes mounted at a first LISE focal point 22.9~m upstream from the final focus, a sequence of four silicon 
detectors (E1: 300$\mu$m, E2: 300$\mu$m, E3: 300$\mu$m, E4: 6mm, respectively) with 
the third one being a silicon-strip detector with 16$\times$16 x-y 
strips with a pitch of 3~mm, and a Germanium array in close geometry. The silicon detectors were 
equipped with two parallel electronic chains with different gains, one for heavy-fragment identification and the 
other for decay spectroscopy.

The fragments of interest were stopped in the third silicon detector of the telescope and identified on an event-
by-event basis by means of their flight times and of their 
energy loss in all detectors of the telescope (see~\cite{blank00ni48,giovinazzo01}). 
Implantation events were triggered by detectors E1 and E2, whereas 
radioactive decay events were triggered either by the detector E3 or by the adjacent silicon detector E4. 
The efficiency to observe a $\beta$ particle in the 
E4 detector for a $\beta$ decay occurring in the implantation detector is about 30\%. 
The $\gamma$ energy calibration and detection efficiency was obtained with standard calibration 
sources. The efficiency was about 1.6\% at 1.3 MeV.

In a 36~h run, 22 $^{45}$Fe implantations were identified. Due to a rather low implantation rate of much less than one radioactive isotope per second in each pixel, 
an implantation-decay correlation could be performed on an event-by-event basis with each 
decay event being correlated to each implantation in the same x-y pixel having occurred less than a well-defined 
maximum time interval before the decay event.

\begin{figure}[htt]
\centerline{
\psfig{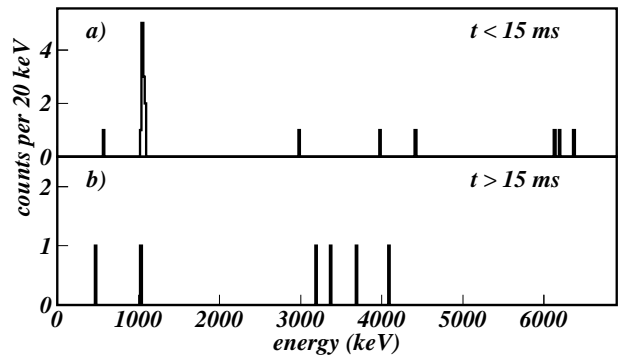}}
\caption{Decay-energy spectrum correlated with $^{45}$Fe implantation. Spectrum a) is obtained under the 
         condition that the radioactive decay occurs faster than 15~ms after implantation, whereas spectrum b)
         contains decay events with decay times between 15~ms and 100~ms. The peak at 1.06~MeV is clearly identified
         as being due to a fast decay.}
\protect\label{fe45_decay}
\end{figure}

\begin{figure}[h]
\centerline{
\psfig{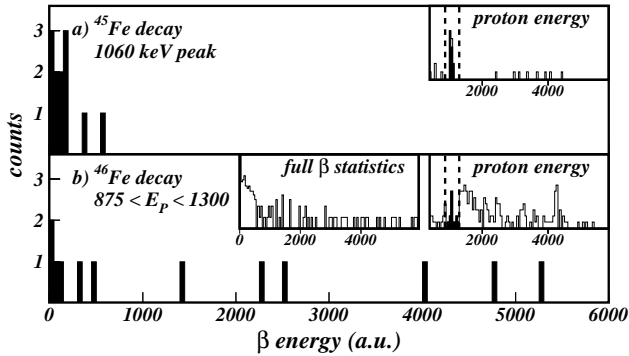}}
\caption{a) $\beta$-particle spectrum from the 6mm-thick detector E4  
         in coincidence with events in the peak at 1.06~MeV. b) Similar
         spectrum obtained after $^{46}$Fe implantation conditioned by a decay-energy range in the E3 detector 
         of $E = 0.875 - 1.3$ MeV, which yields the same statistics  
         as for the 1.06~MeV peak of $^{45}$Fe. The insets show the decay-energy spectrum for the two nuclei 
         with the dark area and the dotted lines indicating the decay-energy condition applied to
         generate the $\beta$ spectra as well as the full-statistics $^{46}$Fe $\beta$ spectrum.}
\protect\label{fe45_beta}
\end{figure}

Figure~\ref{fe45_decay}a shows the decay-energy spectrum correlated with implants of $^{45}$Fe where only decay 
events occurring less that 15~ms after a $^{45}$Fe implantation were analysed. The spectrum exhibits a pronounced 
peak at (1.06$\pm$0.04)~MeV with only a very few other counts. In contrary, in the spectrum in figure~\ref{fe45_decay}b 
conditioned by a decay time in the interval between 15~ms and 100~ms, the 1.06~MeV peak has almost 
completely disappeared and other events higher in decay energy show up. These counts are consistent with the 
decay-energy spectrum of $^{43}$Cr~\cite{giovinazzo01}, the 2p daughter of $^{45}$Fe. The 1.06~MeV peak, however, seems 
to originate only from the fast decay of $^{45}$Fe. In addition, the events in this peak have no coincident 
$\beta$-particle signals in the adjacent detector E4 (see figure~\ref{fe45_beta}a) beyond the noise level, whereas these 
coincident $\beta$ particles can be observed for the $\beta$ decay of $^{46}$Fe implants analysed with a similar condition in energy 
for the E3 detector. For 12 $^{46}$Fe events in the decay-energy interval 0.875~-~1.3~MeV, six 
coincident $\beta$ particles can be identified in the E4 detector (see 
figure~\ref{fe45_beta}b). As the $\beta$-decay end-point energies are roughly the same for $^{45}$Fe and $^{46}$Fe and 
therefore similar $\beta$-detection efficiencies can be assumed, the non-observation of $\beta$ particles in 
coincidence with the 1.06~MeV peak alone is a strong indication that this peak originates from a direct 
two-proton ground-state decay of $^{45}$Fe. The probability to miss all $\beta$ particles for the 12 events in the 
peak is as low as 1.4\%. Although the $\gamma$-detection efficiency was rather low, it is worth mentioning that none of 
the 12 events in the 1.06~MeV peak is in coincidence with a $\gamma$ ray.

The observation of only twelve events in the peak at 1.06~MeV indicates that the branching ratio for 2p decay 
of $^{45}$Fe is not hundred percent, in agreement with theoretical prediction of a $\beta$-decay half-life of about 
7~ms~\cite{ormand96}. However, dead-time losses (the data-acquisition dead time is about 
0.5~ms) make us lose about 3-4 events. The dead zone between two strips 
of the implantation detector is another possible source of losses. The events with an energy above 6~MeV
(see figure~\ref{fe45_decay}a) decay  with a half-life compatible with the one of $^{45}$Fe and are therefore 
must likely $\beta$-delayed events. With this information, a 
branching ratio for 2p emission of 70-80\% can be estimated.

The observation of a pronounced peak at 1.06~MeV from the decay of $^{45}$Fe is 
a strong additional indication that $^{45}$Fe is not decaying by a $\beta$-delayed mode alone as does $^{46}$Fe~\cite{giovinazzo01} with a 
more complexe decay-energy spectrum. In addition, 
the width of the $^{45}$Fe peak (60$\pm$10~keV) is about 40\% smaller than  $\beta$-delayed one-proton peaks from neighboring nuclei.
Although the statsistics is limited, this indicates that the $^{45}$Fe peak is not broadened due to $\beta$ pile-up.

\begin{figure}[h]
\centerline{
\psfig{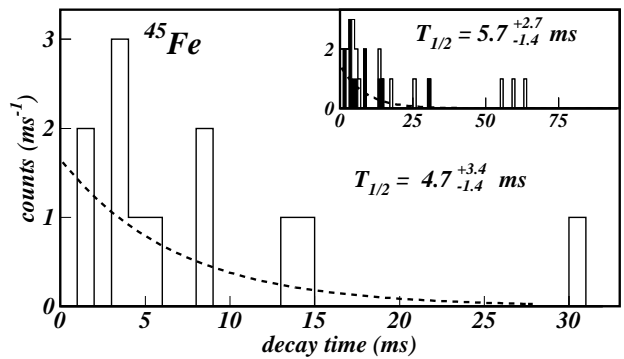}}
\caption{Decay-time spectrum of $^{45}$Fe. The twelve events in the 1.06~MeV peak are fitted by a one-component 
         exponential using the maximum likelihood procedure. The half-life thus obtained 
         is (4.7$^{+3.4}_{-1.4}$)~ms. A fit, including the daughter decay, of all 
         correlated event up to 100~ms after $^{45}$Fe implantation (see inset) yields a half-life of 
         (5.7$^{+2.7}_{-1.4}$)~ms for $^{45}$Fe and a value consistent with literature~\protect\cite{giovinazzo01} 
         for $^{43}$Cr. The dark shaded counts originate from the 1.06~MeV peak.}
\protect\label{fe45_time}
\end{figure}

The decay-time spectrum of $^{45}$Fe gated by the 1.06~MeV peak is shown in figure~\ref{fe45_time}. A one-component 
fit with an exponential yields a half-life for $^{45}$Fe of $T_{1/2}$~= (4.7$^{+3.4}_{-1.4}$)~ms. The decay-time 
spectrum of events up to 100~ms after a $^{45}$Fe implantation can be fitted by taking into account the decay 
of $^{45}$Fe and its 2p daughter $^{43}$Cr. The half-life then is (5.7$^{+2.7}_{-1.4}$)~ms.

\begin{figure}[h]
\centerline{
\psfig{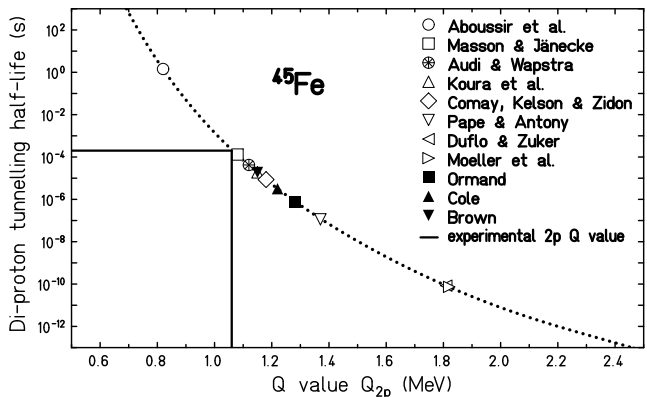}}
\caption{Barrier-penetration half-life as a function of the two-proton Q value, 
         $Q_{2p}$, for $^{45}$Fe. The barrier penetration was calculated by 
         assuming a spectroscopic factor of unity. Different model 
         predictions~\protect\cite{brown91,ormand96,cole96,masses88,moeller95,duflo95,aboussir95,audi97,koura00}
         were used for $Q_{2p}$. The experimentally observed Q value of $^{45}$Fe 
         implies a di-proton barrier-penetration half-life of 0.2~ms.}
\protect\label{fe45_qvalue}
\end{figure}

The energy of the peak at 1.06~MeV agrees nicely with $Q_{2p}$ value predictions from 
Brown~\cite{brown91} of 1.15(9)~MeV, from Ormand~\cite{ormand96} of 1.28(18)~MeV, and 
from Cole~\cite{cole96} of 1.22(5)~MeV. These models use the isobaric-multiplet mass equation 
and shell-model calculations to determine masses of proton-rich nuclei from their neutron-rich mirror 
partners. Q-value predictions from models aiming at predicting masses for the entire chart of nuclei are in less 
good agreement with our present result. All the 2p Q-value predictions are summarized in figure~\ref{fe45_qvalue}, 
where they are used in barrier-penetration calculations using the simple di-proton model for two-proton 
emission (the model used in \cite{brown91} with $R=4.2$ fm).

For $E_p$~= 1.06$\pm$0.04~MeV, these calculations predict a di-proton barrier penetration half-life of (0.2$^{+0.3}_{-0.1}$)~ms, 
if one assumes a spectroscopic factor of unity. Brown~\cite{brown91,brown91a} calculated a 
spectroscopic factor of 0.195 for a direct 2p decay of $^{45}$Fe which increases 
the barrier tunnel time to a predicted half-life of (1.2$^{+1.5}_{-0.5}$)~ms. 

We have also used the R-matrix formulation of Barker~\cite{barker01} both in its simple form, equivalent to the di-proton decay
model discussed above, and in a more realistic form which includes the s-wave resonance of the two protons.
In the R-matrix we need to choose a channel radius $a$ and an associated dimensionless reduced width
$\theta^2_{sp}$ (Eq. 16 of \cite{barker99}). For consistency we take $a=4.2$ fm which is a region of the resonance
wave function where
$\theta^2_{sp} \approx 1$. With $S=0.195$ we obtain a half-life of (1.7$^{+2.3}_{-0.5}$)~ms in the 
simple model, and (8$^{+22}_{-6}$)~s in the s-wave resonance model. The reduction in the phase space 
due to the two-proton resonance increases the lifetime by a factor of 4000.
Grigorenko et al.~\cite{grigorenko00} studied the emission of two protons from $^{48}$Ni in a three-body model. 
They find that the half-life for the same $Q_{2p}$ value increases by about three orders of magnitude 
when going from a simple di-proton model to the more advanced three-body model. Thus the models which include the
interaction between two-protons appear to give a lifetime which is much longer than the observed value.

A sequential decay seems to be excluded for $^{45}$Fe, as the intermediate state, the ground state of $^{44}$Mn, is, 
depending on the prediction used, either not in the allowed region or at the very limit of the allowed region. The 
model predictions~\cite{brown91,ormand96,cole96} range from $Q_{1p}$~= -24~keV to +10~keV. As the 
intermediate state is most probably rather narrow (barrier-penetration calculations yield a value of about 
$\Gamma$~= 50~meV), the first proton would have a very long half-life (hours or even days), which makes this 
decay mode very unlikely.

In conclusion, we have studied the decay of $^{45}$Fe in an experiment where we measured its decay energy yielding 
a peak at 1.06~MeV, a half-life of 4.7~ms, and, in particular, no coincident $\beta$ particle or $\gamma$ ray for 
the events in the decay-energy peak. In addition, strong indications for the decay of the 2p daughter, $^{43}$Cr, after 
implantation of $^{45}$Fe are found. Additional support comes from the width of the $^{45}$Fe peak.
The energy of the observed peak is in reasonable agreement with theoretical predictions for the 2p 
decay energy. A consistent picture arises, if one assumes that a two-proton ground-state emission occurs. 
Simple model calculations favor the observed decay mode to be a $^{2}$He/di-proton emission.

Whereas the two-proton ground-state decay of $^{45}$Fe seems to be established with the present data, future 
high-statistics data should definitively allow to conclude on the nature of the two-proton decay, $^{2}$He emission 
or three-body decay. This question can be addressed in an experiment, which measures the individual proton energies 
and the relative-angle distribution for the two protons emitted which should be either isotropic (three-body decay) 
or forward-peaked ($^{2}$He emission). 

It should be mentioned that similar results, although with less statistics and a lower energy resolution, were also obtained 
at the FRS of GSI~\cite{pfuetzner02}.

We would like to acknowledge the continous effort of the whole GANIL staff for ensuring 
a smooth running of the experiment. This work was 
supported in part by the Polish Committee of Scientific Research under
grant KBN 2 P03B 036 15, the 
contract between IN2P3 and Poland, the NSF grant PHY-00-7911, as well as by the Conseil R\'egional 
d'Aquitaine.

\end{document}